\documentclass[reprint,superscriptaddress,prb,showkeys]{revtex4-2}
\usepackage{empheq}
\usepackage{float}
\usepackage{lipsum}
\usepackage{mathtools, cuted}
\usepackage{esvect}
\usepackage{xcolor}
\usepackage[caption = false]{subfig}
\usepackage{graphicx}
\usepackage{dcolumn}
\usepackage{bm}
\usepackage{hyperref}
\usepackage{physics}
\usepackage[mathlines]{lineno}
\usepackage[capitalise]{cleveref}
\usepackage[english]{babel}
\usepackage{orcidlink}
\usepackage{amsmath}
\usepackage{amssymb}
\usepackage{comment}
\usepackage{listings}

\usepackage{algorithm}
\usepackage{algpseudocode}

\newcommand{\brac}[1]{\left(#1 \right)} %nice brackets

\begin{document}

%\title{Multistimulus Machine Learning in Microfluidic Memristor Networks}
\title{Multimodal Physical Learning in Brain-Inspired Iontronic Networks}
\author{Monica Conte}
\affiliation{Soft Condensed Matter \& Biophysics, Debye Institute for Nanomaterials Science, Utrecht University, Princetonplein 1, 3584 CC Utrecht, The Netherlands,}
\author{René van Roij}
\affiliation{Institute for Theoretical Physics, Utrecht University, Princetonplein 5, 3584 CC Utrecht, The Netherlands.}
\author{Marjolein Dijkstra}
\affiliation{Soft Condensed Matter \& Biophysics, Debye Institute for Nanomaterials Science, Utrecht University, Princetonplein 1, 3584 CC Utrecht, The Netherlands,}

\date{\today}

\begin{abstract}
Inspired by the brain, we present a physical alternative to traditional digital neural networks---a microfluidic  network in which  nodes are connected by conical, electrolyte-filled channels acting as memristive iontronic synapses. Their  electrical conductance responds not only to electrical signals, but also to chemical, mechanical, and geometric changes. Leveraging this multimodal responsiveness, we develop a  training algorithm where learning is achieved by altering either the channel geometry  or the applied  stimuli. The network performs  forward passes  physically via ionic relaxation, while learning combines this physical evolution with numerical gradient descent.  We theoretically demonstrate that this system can perform  tasks like  input-output mapping and linear regression with bias, paving the way for soft, adaptive materials that  compute and learn without  conventional electronics.
\end{abstract}
\maketitle

\renewcommand{\figurename}{FIG.}

Modern computing systems excel at  performing complex numerical and symbolic operations with high precision. However, they still lag behind biological brains in  perceptual tasks such as pattern recognition, image interpretation, and language understanding \cite{lake2017building}. Unlike traditional computers, the brain's ability to process information through massively parallel and distributed networks has inspired the development of artificial neural networks (ANNs). These models are at the core of  many recent advances in  artificial intelligence (AI) \cite{ai_tech1} with applications ranging from image and speech recognition~\cite{pattern_recognition,review_image_rec, speech_rec} to natural language processing\cite{naturallang}. Recent breakthroughs in AI have mainly been driven  by advances in  digital hardware---particularly Graphics Processing Units---and the availability of big data. However, the growing energy demands and hardware limitations of modern AI systems are becoming increasingly problematic \cite{energy_cost_ml1, energy_cost_ml2}. 
%As ANNs scale in size and complexity, their energy consumption rises sharply. 
At the same time, the progress in  silicon-based hardware---once predicted by Moore’s Law---is slowing down, creating an increasing mismatch between computational demand on the one hand and hardware capability and energy consumption on the other hand. This has motivated the search for alternative computing paradigms that are both energy-efficient and scalable. 

One promising direction is the development of physical neural networks (PNNs), which perform learning and inference using physical processes rather than  digital computations   \cite{trainingphysicalneuralnetworks,backporp_pnn, sternbecomelearn, physicalconsinmemroy,falk2025temporal,stern2023learning,stern2024physical,anisetti2023learning}. Interestingly, many physical systems can perform specific tasks  %faster and 
more energy-efficiently  than conventional electronics \cite{ai_energy_consumpt}, making PNNs a compelling route toward more sustainable and scalable AI. Like ANNs, PNNs learn by adjusting internal parameters---called learning degrees of freedom or weigths and biases in machine learning---to achieve a desired input-output relationship. However, the adjustments in PNNs emerge from a direct physical response to external stimuli rather than from software-based updates in ANNs. By now, PNNs have already been explored across optical \cite{optics1},  electronic \cite{electronics1}, photonic \cite{acoustic1}, and mechanical platforms \cite{mechanics1}. However, most implementations rely on a single physical mechanism to update the trainable parameters, which limits both the flexibility of training protocols and the  tunability of the system.  Moreover, many  training algorithms rely entirely on {\em in-silico} optimization, complicating transfers of trained models to real-world physical systems \cite{electronics1, insilicottrain2}. To overcome this challenge, hybrid training approaches have been proposed in which the forward (inference) pass is executed by the physical network itself, while the backward pass is computed numerically on an external computer \cite{backporp_pnn}. Another strategy, known as coupled learning \cite{supervised_learning}, uses  local learning rules applied directly at the level of individual network components  \cite{demonstr_coupled_learning, sternquasistatic, sterncoupled, chemical}. Although these local rules eliminate the need to optimize a global cost function, as required in standard backpropagation, they  often require two identical copies of the physical system to apply clamped and free boundary conditions \cite{demonstr_coupled_learning} which imposes practical limitations in many experimental setups.

\begin{figure*}[t]
    \centering
    \includegraphics[width=1.9\columnwidth]{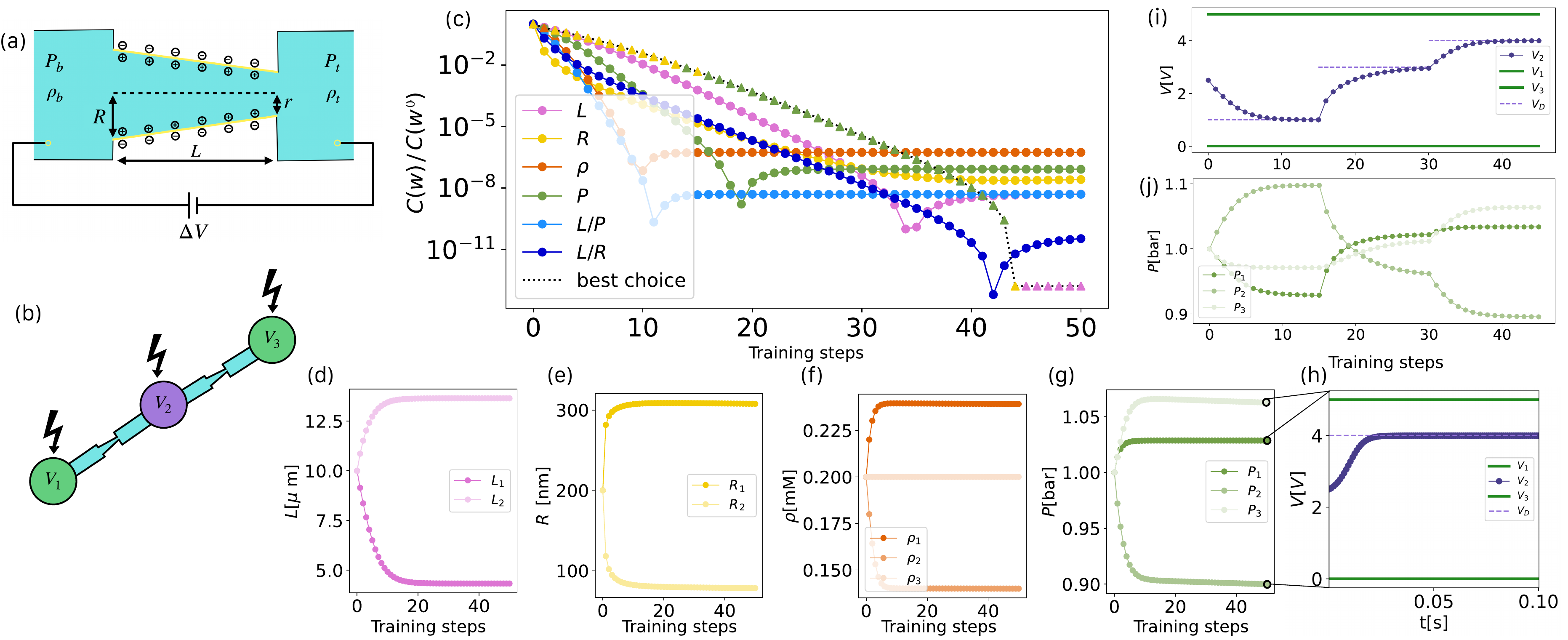}
    \caption{(a) Illustration of a microfluidic memristor channel with applied external potential. (b) Schematic of a memristor-voltage divider. (c)  Relative cost function $C(\boldsymbol{w}^s)/C(\boldsymbol{w}^0)$ as a function of the training steps, obtained by training with (d) lengths (pink), (e) concentrations (orange), (f) base radii (yellow),  (g) pressures (green), with learning rates $\alpha = [3 \cdot 10^{-6}, 3 \cdot 10^{-6}, 5 \cdot 10^{-4}, 100]$, respectively, combined weight subsets: length and pressure (light blue), length and base radius (dark blue), as well as for the adaptive ``best weight'' choice (triangles, connected by a black dotted line). (h) Time evolution of $V_2(t)$ for the pressure-trained network, after switching on the input $V_1= 5 V$ at $t=0$. The evolution during the pressure-training step $s$ of (i) the steady-state output voltage $V_2({\bf w}^s)$ (symbols) towards three different consecutive desired voltages $V_2^D$ (horizontal dashed lines), with (j) the associated re-trained pressures, starting at $s=0$ from the default parameter set. }
    \label{fig:mse_weights_vd}
\end{figure*}

In this Letter, we study iontronic PNNs that more closely emulate the multimodal responsiveness of biological neuronal systems  than conventional digital architectures. For instance, the brain excels at simultaneously processing diverse  stimuli---electrical, chemical, and mechanical---a capability known as multimodal responsiveness. This feature is believed to play a central role in the brain’s remarkable efficiency, adaptability, and robustness \cite{deneve2017brain,balasubramanian2021brain}.

We present networks of aqueous 1:1 electrolyte reservoirs, which play the role of nodes, connected by cone-shaped microfluidic channels that act as edges, as shown in \cref{fig:mse_weights_vd}(a,b). Each channel allows for the transport of water and salt ions between the two connected reservoirs. Our focus is solely  on the voltage-driven electric current carried by the ions, i.e. we treat the network as an electric circuit. As in standard Ohmic conductors the electrical conductance of a cone-shaped channel depends on its geometric parameters, such as length, base radius, and tip radius. Interestingly, however, the electrical conductance of a  microfluidic conical channel also depends on the pressures and salt concentrations in the  connected reservoirs, and can vary over time if the voltage drop is time-dependent \cite{Rene_PressureSensitive,Rene_IonicNeuromorphic}. This dependence on pressure, salt concentration, and time is caused by electrokinetic effects such as streaming, electro-osmosis, and diffusio-osmosis, which occur on timescales ranging from milliseconds to seconds for ions in micron-sized systems. We exploit these dependencies on channel geometry and external stimuli by using them as a versatile set of weights to enable iontronic PNNs to learn desired electric input-output relations and to perform computational tasks.  

We consider a PNN composed of $N$ nodes $i=1,\dots,N$ connected by $M$ edges between selected pairs of nodes $\langle i,j\rangle$. The PNN represents an electric circuit with (possibly time-dependent) voltages $V_i(t)$ and conductances $g_{ij}(t)$ on the nodes and  edges, respectively, such that an electric current 
$I_{ij}(t)=g_{ij}(t)(V_i(t)-V_j(t))$ can flow between the connected pairs, where we understand that $g_{ij}(t)\equiv 0$ if $i$ and $j$ are not connected. The set of nodes is divided into three subsets ${\cal I}$, ${\cal H}$, and ${\cal O}$, representing input nodes (green), hidden nodes (grey), and output nodes (purple), with   colors referring  to those in all figures. At least two input nodes are required, of which one is grounded for convenience. The input nodes $i\in {\cal I}$ have known potentials  $V_i(t)$ and  inject or extract  currents $I_i(t)$, which then propagate  through the network. In absence of capacitive and inductive elements, the Kirchhoff equations $\sum_k I_{kj}=0$ must hold for all  hidden and output nodes  $j\in{\cal H}\cup{\cal O}$. The Kirchhoff equations reduce to a linear-algebra problem for the $N$ unknowns $I_i(t)$ with $i\in{\cal I}$ and $V_j(t)$ with $j\in{\cal H}\cup{\cal O}$, provided $g_{ij}(t)$ is known.

Each node of our iontronic PNN hosts an aqueous 1:1 electrolyte reservoir at room temperature  endowed with a voltage $V_i(t)$, a static pressure 
$P_i$, and a static salt concentration 
$\rho_i$.
The microfluidic channels connecting selected node pairs $\langle i,j\rangle$ have azimuthal symmetry, length $L_{ij}$, base radius $R_{ij}$, and tip radius $r_{ij}$. We present a schematic representation of a single channel in \cref{fig:mse_weights_vd}(a). The combination of  cone-shaped geometry $r_{ij}<R_{ij}$ and a nonzero surface charge on the channel walls has been shown \cite{Rene_IonicNeuromorphic} to make the {\em steady-state} electric conductance $g_{\infty,{ij}}(V_i-V_j;\boldsymbol{W}_{ij})$ voltage-dependent. Here we also make explicit its dependence on the set of weights $\boldsymbol{W}_{ij}=(P_i,\rho_i,L_{ij}, R_{ij},r_{ij},P_j,\rho_j)$, see End Matter for explicit expressions. For a time-dependent voltage drop $V_i(t)-V_j(t)$ across  channel $\langle i,j\rangle$, the time-dependent conductance $g_{ij}(t)$  is well described by  
\begin{equation}
\frac{\partial g_{ij}(t)}{\partial t} = \frac{g_{\infty,ij}\brac{V_i(t)-V_j(t);\boldsymbol{W}_{ij}}-g_{ij}\brac{t}}{\tau_{ij}},
\label{eq:eqofmotconductance}
\end{equation}
where the memory retention time $\tau_{ij}=L_{ij}^2/12D$ arises from the finite ion diffusion coefficient $D=1.75 ~\mu \text{m}^2 \text{ms}^{-1}$  \cite{Rene_IonicNeuromorphic}. Note that $\boldsymbol{W}_{ij}$ is assumed to be static. Thus, rather than solving  the elementary linear-algebra problem of the Kirchhoff equations for $N$ unknowns at  given $g_{ij}(t)$, we are now confronted with solving a closed set of $N$ nonlinear algebraic Kirchhoff equations coupled to $M$ first-order differential equations (\ref{eq:eqofmotconductance}) for the $M$ conductances $g_{ij}(t)$. This yields a total of $N+M$ unknowns to be determined at fixed $\boldsymbol{W}_{ij}$ and  initial conditions $g_{ij}(t=0)$. Numerically, however, this remains  rather straightforward for the relatively small networks of interest here.

For a given network, we define for convenience a default parameter set $\boldsymbol{W}$ of initial weights, consisting of $2N+3M$ elements given by on-site pressures $P_i=1\text{~bar}$ and salt concentrations $\rho_i=0.2 \text{~mM}$ for all nodes combined with  on-edge lengths $L_{ij}=10~\mu\text{m}$, base radii $R_{ij}=200\text{~nm}$, and tip radii $r_{ij}=50\text{~nm}$ for all channels. With these defaults, the memory retention time is  $\tau_{ij}=4.8\text{~ms}$. After switching on static input voltages $V_i$ for $i\in{\cal I}$, the network relaxes over several (tens of) milliseconds  to a steady-state set of static output potentials $V_j$ for $j\in\cal O$.  This allows us  to train the network to produce a desired input-output relation, in this case a mapping of the input potentials $V_i$ for $i\in{\cal I}$ to the desired output potentials $V_j^D$ for $j\in{\cal O}$, by adjusting a subset $\boldsymbol{w}\subset \boldsymbol{W}$ of  system parameters while holding the others fixed at their default values. For example, $\boldsymbol{w}$ could be just the on-site pressures ($\boldsymbol{w}=\{P_1,\cdots P_N\}$) or only the on-edge lengths ($\boldsymbol{w}=\{L_1,\cdots,L_M\}$). 
We define the \emph{cost function}
\begin{equation}
C(\boldsymbol{w}) = \sum_{j \in {\cal O}} \brac{V_j\brac{\boldsymbol{w}} - V_j^D}^2,
\label{eq:cost_func}
\end{equation}
where $V_j(\boldsymbol{w})$ denotes the (time-relaxed) potential at output node $j$ with the variational parameter set $\boldsymbol{w}$ and $V_j^D$ is the desired output voltage.  The parameter set $\boldsymbol{w}$ is updated from training step $s=0,1,2,\dots$ to  $s+1$ using the standard steepest descent method, such that 
\begin{equation}
\boldsymbol{w}_m^{s+1} = \boldsymbol{w}_m^{s} - \alpha\frac{\Delta C}{\Delta w_m}, 
\label{eq:weight_update}
\end{equation}
where $\alpha>0$ is an adaptable learning rate, that we make dimensionless by the rescaling  $\alpha C(\boldsymbol{w}^{0})/({\boldsymbol{w}^0})^2\rightarrow \alpha$. 
In Eq.(\ref{eq:weight_update}), $\Delta C/\Delta w_m$ is a first-order forward finite difference approximation to the partial derivative $\partial C(\boldsymbol{w})/\partial w_m$ calculated using a small step size $\Delta w_m=10^{-3}w_m^0$, where $w^{s}_m$ is the $m$-th component of $\boldsymbol{w}$ at training step  $s$. 
During each training step $s$, the voltages  $V_j(t)$ and conductances $g_{ij}(t)$ relax  physically over timescales of (tens of) milliseconds, first at fixed weights $\boldsymbol{w}^s$ to calculate $C$ and then at shifted weights to calculate the $\Delta C$'s. Next, the weights are updated from  $\boldsymbol{w}^s$ to $\boldsymbol{w}^{s+1}$ via the steepest descent of Eq.(\ref{eq:weight_update}). This cycle of physical relaxation and weight updating is repeated until  $C$ is minimized. We note that this iterative  process does not rely on backpropagation as the gradient components $\partial C/\partial w_m$ are obtained from two physical processes at weights that differ by a small $\Delta w_m$.

To demonstrate the multi-modal learning capabilities of these microfluidic channels, we first consider a simple voltage divider, which is the smallest possible electric circuit consisting of  $N=3$  nodes connected in series by $M=2$ channels, as illustrated in Fig. \ref{fig:mse_weights_vd}(b). The two external nodes are fixed at input voltages $V_1=5\text{~V}$ and $V_3=0\text{~V}$, and the desired voltage at the central node is set to $V_2^D=4\text{~V}$. Starting from our default parameter set $\boldsymbol{W}$, we train the network for 50 steps  by optimizing four different subsets $\boldsymbol{w}$ of weights separately. The  evolution of the cost function $C(\boldsymbol{w}^s)$, normalized by its initial value $C(\boldsymbol{w}^0)$, is shown in Fig.\ref{fig:mse_weights_vd}(c)  as a function of training steps. It reveals a reduction of $C$ by 6 to 9  orders of magnitude within $10$ to $40$ steps for all four cases, demonstrating  that each of these physical parameters can efficiently contribute to learning. We also plot the evolution of the four sets of weights during training in Fig.\ref{fig:mse_weights_vd}(d)-(g). 
For the final weights $\boldsymbol{w}^{50}$ obtained in (g) we further show, in \cref{fig:mse_weights_vd}(h), the autonomous physical time evolution of $V_2(t)$, after switching on the input $V_1=5\text{~V}$. The voltage $V_2(t)$ approaches its target value  $V_2^D$ within about $20\text{~ms}$, consistent with the characteristic timescale $\tau_{ij}$ for these parameters.

Motivated by the intuition that increasing the number of  degrees of freedom may enhance the learning capacity, we explore the simultaneous  training of multiple weight subsets. We begin by  combining different geometrical design parameters by selecting $\boldsymbol{w} = \{L_1, \dots, L_M\} \cup \{R_{1}, \dots, R_{M}\}$, with the resulting cost function shown in \cref{fig:mse_weights_vd}(c). Indeed, combining multiple geometrical properties significantly enhances training performance, improving accuracy by up to  three orders of magnitude. However, when combining node- and edge-defined properties, for example, $\boldsymbol{w} = \{L_1, \dots, L_M\} \cup \{P_{1}, \dots, P_{N}\}$, the accuracy of the cost function does not surpass that of the lengths-only case.
Finally, we implemented a training procedure in which  the network autonomously selects the weight update that yields the greatest reduction in the cost function. This result, shown in \cref{fig:mse_weights_vd}(c) as a black dotted line with triangles colored according to the selected  weight, demonstrating that this adaptive update strategy achieves  the lowest $C$ and therefore  the best overall training performance.

To examine the versatility and adaptability of the algorithm, we also pressure-trained the voltage divider for three consecutive target voltages $V_2^D=\{1V, 3V, 4V\}$.  In \cref{fig:mse_weights_vd}(i)  the dashed lines indicate the target values  $V_2^D$  and their abrupt  changes at training steps $s=0$, 15, and 30. The dots denote  the time-relaxed voltages $V_2(\boldsymbol{w}^s)$ at each step $s$, which converge to the new target $V_2^D$ within roughly 10 steps. The corresponding evolution of the learned weights, in this case the pressures $P_i$ at the three nodes, is shown at each $s$ in \cref{fig:mse_weights_vd}(j). These results demonstrate that the iontronic channels can be retrained on demand at any stage to achieve new  objectives, highlighting their reconfigurability and  reusability.  
 \begin{figure*}[t]
    \centering
    \includegraphics[width=0.7\textwidth]{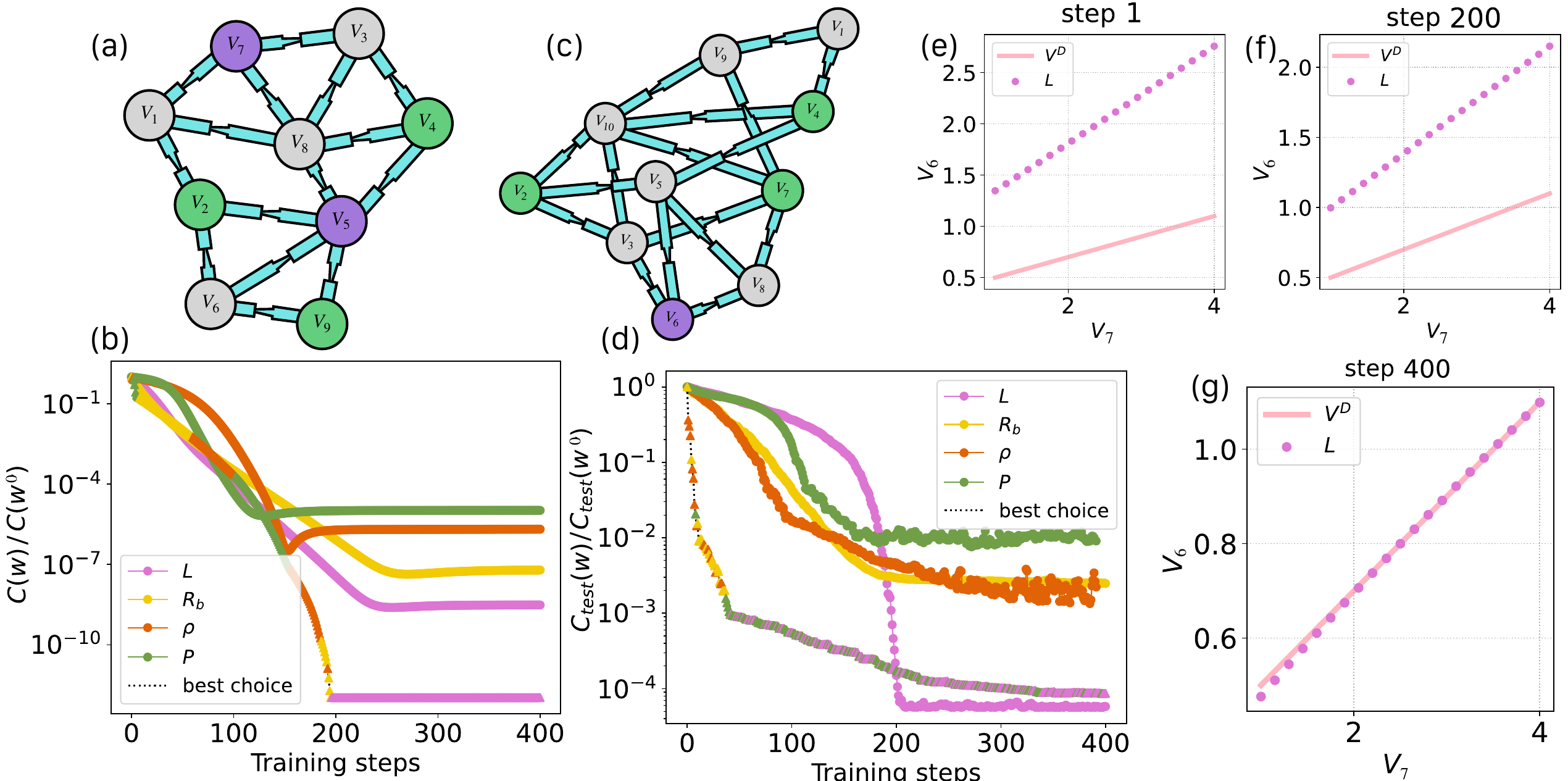}
    \caption{(a) Generic network of $9$ nodes and $15$ edges. (b) Relative cost function for five subsets of weights indicated by colors, for learning rates $\alpha=[1\cdot 10^{-6}, 8\cdot 10^{-7}, 1\cdot 10^{-4}, 20]$.  (c) Network geometry used for linear regression training. Among the three input nodes, one is used to ground the circuit ($V_2 = 0V$), one ($V_4$) provides a constant voltage source for bias, and the remaining one ($V_7$) is used to input the training data point. For the four weight types (channel length, base radius, ion concentration  and pressure), we used as constant voltage input $V_4 = [11V,4V,4V,11V]$ and learning rates $[2\cdot 10^{-7}, 1\cdot 10^{-6}, 1\cdot 10^{-4}, 2\cdot 10^{2}]$. (c) Test cost function evaluated every $10$ training steps for four different weight choices. Figures (e), (f) and (g) show visual representations of the network’s output trained with varying channel lengths at three different training steps. }\label{fig:mse_general_regression}
\end{figure*} 

Next, we generalize the training objective and expand the iontronic circuit to $N=9$ nodes connected by $M=15$ cone-shaped microfluidic channels in the more complex geometry of \cref{fig:mse_general_regression}(a). This network features three input nodes with fixed voltages  $(V_2,V_4,V_9)=(0,2,5)\text{~V}$, four hidden nodes, and two output nodes with target  voltages $(V_5^D,V_7^D)=(3,4)\text{~V}$. As before, training is performed on a predefined subset of weights $\boldsymbol{w}$ containing either the $N$ pressures, the $N$ concentrations, the $M$ lengths,  or the $M$ base radii, starting from the default parameter set $\boldsymbol{W}$. 
Successfully realizing such a prescribed  input-output voltage mapping in a generic electric circuit demonstrates the versatility of our approach and its potential for a broad range of  electronic applications  \cite{voltage_controlled_oscillator, onchip_powerdistributed}.

The evolution of the cost function $C(\boldsymbol{w}^s)$ over 400 training steps is shown in \cref{fig:mse_general_regression}(b) for each of the four subsets of weights. In all cases,  the cost function decreases  by several orders of magnitude, with the largest reduction obtained by optimizing the channel lengths. Consistent with the voltage divider results, the adaptive ``best choice'' update rule for the weights achieves the most efficient learning, outperforming all other training protocols by up to three orders of magnitude. 

Finally, we investigate to what extent the network in \cref{fig:mse_general_regression}(c), consisting of $N=10$ nodes connected by $M=17$ microfluidic iontronic channels, can perform the elementary yet  nonlinear regression task defined by the target relation for the output voltage $V_6^D=aV_7+b$ at node 6, across the full range of input voltages $V_7\in[1,4]\text{~V}$. To be specific, we set  $a=0.2$ and $b=0.3\text{~V}$.  
Training starts from the default parameter set $\boldsymbol{W}$, selecting one of the four subsets $\boldsymbol{w}$ of weights as before, either  pressures (with $V_4=11\text{~V}$), concentrations (with $V_4=4\text{~V}$), lengths (with $V_4=11\text{~V}$), or base radii (with $V_4=4\text{~V}$). At each training step $s=0,\dots,400$, a random input voltage $V_7\in[1,4]\text{~V}$ is drawn uniformly, and  the corresponding time-relaxed $V_6$ is determined from solving the Kirchhoff equations extended with Eq.(\ref{eq:eqofmotconductance}). The cost function $C(\boldsymbol{w}^s)$ and its change $\Delta C$ are then evaluated via  Eq.(\ref{eq:cost_func}), followed by the weight update  $\boldsymbol{w}^{s+1}$. To assess the learning progress, we compute for each of 20 inputs $V_7^{(k)}\in[1,4]\text{~V}$ ($k=1,\cdots,20$) the individual cost  $C_k^s\equiv C(\boldsymbol{w}^s)$,  and define the total test cost function $C_{\text{test}}(s)\equiv\sum_{k=1}^{20} C_k^s$. The performance of the iontronic network on this task is shown in \cref{fig:mse_general_regression}(d), where learning rates were optimized to balance stability and convergence speed.  The test cost function  $C_{\text{test}}(s)$ decreases to about $10^{-4}$ for length-based weights and to below   $10^{-2}$ for the other three weight sets, demonstrating the circuit's ability  to learn a nonlinear mapping by tuning its physical  parameters. The higher driving voltage required for length and concentration training ($11V$) compared to pressure- and radius-based training ($4V$) probably reflects differences in their underlying ionic transport mechanisms. 
A visual comparison of the desired and length-learned input-output relations of the trained network is presented in \cref{fig:mse_general_regression}(e), (f) and (g). 

In conclusion, we have demonstrated that networks  of cone-shaped microfluidic channels function as  highly  versatile iontronic circuits capable of dynamically (re-)learning complex electrical input-output relationships through the precise tuning of multiple  physical weights, most notably pressures, salt concentrations, and channel geometries. The choice of weight type is flexible and can be tailored to specific applications. We anticipate that these results will strongly stimulate   experimental realizations and practical applications of such adaptive iontronic devices. Moreover, the trainability and functionality of these devices could be significantly exhanced by incorporating non-electrical control mechanisms, such as ion release from responsive hydrogels or shape morphing via  liquid crystal elastomer channels, paving the way  for a new generation of adaptive soft electronics and intelligent iontronic circuits.  

{\bf Acknowledgement:} We thank Nex Stuhlm\"{u}ller and David Santiago Quevedo for fruitful discussions, Tim Kamsma for providing an explicit expression of the static channel conductance in the case of a concentration drop.

\clearpage
\bibliographystyle{unsrt} 
\bibliography{biblio}

\clearpage

\begin{center}
\begin{widetext}
\textbf{End Matter}        
\end{widetext}
\end{center}
\appendix

\subsection{Steady-state and time-dependent conductance of a cone-shaped channel}
\label{app:cone}
Let us follow Refs.\cite{Rene_PressureSensitive,Rene_IonicNeuromorphic} and consider an azimuthally symmetric conical channel that connects two reservoirs of an incompressible aqueous 1:1 electrolyte with viscosity $\eta=1.0$  mPa s and electric permittivity $\epsilon = 0.71$ nF $\text{m}^{-1}$ containing ions with diffusion coefficient $D = 1.75 ~\mu \text{m}^2 \text{ms}^{-1}$ and charge $\pm e$, with $e$ the proton charge. The cone has base radius $R = 200$ nm, tip radius $r=50$ nm, and length $L = 10~\mu$m, which is in the long-channel limit where entrance and exit effects can be neglected. The walls of the channel are homogeneously charged with a surface charge density $e\sigma$, which generates a nonzero surface potential $\psi_0$ 
that attracts (oppositely-charged) counterions and repels (like-charged) coions, such that an electric double layer (EDL) is formed that screens the surface charge.  Due to this screening cloud the electric potential decreases exponentially with distance from the surface of the channel with a decay length equal to the Debye length, which depends on the salt concentration and is typically in the 1-10 nm regime for the parameters of our interest.

At the far side of the reservoir at the base and tip side of the channel, the electric potentials $V_b$ and $V_t$, the pressures $P_b$ and $P_t$, and the (bulk) ion concentrations $\rho_b$ and $\rho_t$ are imposed, respectively. Thus, if the microfluidic channel is exposed to a nonzero potential drop $\Delta V = V_b - V_t$, 
an electric current is generated due to migration of mobile ions of the electrolyte. Likewise, a non-zero pressure drop $\Delta P = P_b - P_t$ induces a Poiseuille-like flow $Q_P = (\pi R^3 r^3/8L\eta\langle R^2\rangle)\Delta P$.  Moreover, the potential drop also generates
an electro-osmotic fluid flow $Q_V = (\pi R r\epsilon\psi_0/L\eta)\Delta V$, and in principle a nonzero salt concentration difference $\Delta \rho = \rho_b - \rho_t$ 
generates a diffusio-osmotic fluid flow, however this contribution to the total flow $Q$ is negligible for the parameters used in this work. Thus, we consider only the pressure and potential contributions to the total fluid flow, $Q = Q_P + Q_V$, which we characterise below by the dimensionless P\'{e}clet number
$\text{Pe} = QL/D\pi r^2$ that quantifies the importance of advection over diffusion \cite{Rene_PressureSensitive}. 
%assuming the diffusive flow to be negligible.  
The applied voltage and the resulting fluid flow are known to generate non-trivial ion concentration profiles in a cone-shaped channel, which, in turn, directly influence the channel conductance $g$ \cite{Rene_PressureSensitive,Rene_IonicNeuromorphic}.

In the steady-state, a cone-shaped channel of fixed geometry is considered to be exposed to a constant driving by $\Delta V$, $\Delta P$, and/or $\Delta\rho$. For later convenience we introduce the short-hand notation $\boldsymbol{W}=(P_b,P_t,\rho_b,\rho_t,L,R,r)$, where we note that the pressures and salt concentrations are ``node parameters'' while the geometric parameters are defined on the edges.   We can thus denote the corresponding steady-state conductance by $g_{\infty}\brac{\Delta V,\boldsymbol{W}}$. We build on Ref.\cite{Rene_IonicNeuromorphic} and write
\begin{equation}
\frac{g_{\infty}\brac{\Delta V,\boldsymbol{W}}}{g_0} = \int_0^{L} \frac{\bar{\rho}\left(x;\Delta V, \boldsymbol{W}\right)}{2\rho_b L} dx  \label{ginf}                                         \end{equation}
where $g_0=(\pi R r/L)(2\rho_b e^2D/k_B T)$ is the Ohmic conductance at vanishing driving and $\bar{\rho}(x;\Delta V,\boldsymbol{W})$ the radially averaged salt concentration as a function of $x\in[0,L]$ \cite{Rene_IonicNeuromorphic}. By analytically solving the stationary condition of the total salt flux in the thin-EDL and long-channel limit, an expression for the salt concentration profile can be derived \cite{Rene_PressureSensitive}. It reads
\begin{multline}
\frac{\bar{\rho}\left(x;\Delta V, \boldsymbol{W}\right)}{2\rho_b} = 1 - \frac{\Delta \rho}{\rho_b} \mathcal{I}\left(x; \Delta V, \boldsymbol{W}\right) \\
 + \frac{ \rho_{in}\left( \Delta V,\boldsymbol{W} \right) }{2\rho_b \text{Pe}} \left[ \mathcal{F}(x) - \mathcal{I}\left(x; \Delta V, \boldsymbol{W}\right) \right],
\label{eq:rho_average}
\end{multline}
where we defined the two functions
\[
\mathcal{I}\left(x; \Delta V, \boldsymbol{W}\right) = \frac{e^{\frac{x}{L}\frac{r^2}{R R(x)} \text{Pe}}-1}{e^{\frac{r}{R} \text{Pe}}-1}, \ \ \ \ \mathcal{F}(x;\boldsymbol{W}) = \frac{x}{L}\frac{r}{R(x)},
\]
and the concentration inhomogeneity parameter
\[
\rho_{in}\brac{\Delta V,\boldsymbol{W}} = 2\frac{(R-r)\sigma}{r^2}\frac{e\Delta V}{k_B T}.
\]
The channel radius at axial position $x\in[0,L]$ is written as $R(x)=R - (x/L)(R-r)$. Here one should also realise that the Péclet number depends on $\Delta V$ and $\boldsymbol{W}$. Thus, the static conductance $g_\infty(\Delta V, \boldsymbol{W})$ follows from a straightforward numerical evaluation of Eq.(\ref{ginf}). 

When the external stimulus $\Delta V(t)$ is not steady but varies with time $t$,  the total salt concentration profile and hence the conductance will also change with time. However, it takes time for the ions to get transported into or out of the channel to form the concentration profile $\bar{\rho}(x)$, and as a consequence the conductance $g(t;\Delta V(t),\boldsymbol{W})$ at time $t$ is not equal to the instantaneous static conductance $g_\infty(\Delta V(t),\boldsymbol{W})$.  Here we assume throughout for convenience that $\boldsymbol{W}$ is time-independent, although this assumption can easily be relaxed. We follow Ref.\cite{Rene_IonicNeuromorphic} and consider the case that $g(t;\Delta V(t),\boldsymbol{W})$ relaxes 
towards $g_{\infty}(\Delta V(t),\boldsymbol{W})$ on a diffusion-like time scale $\tau = L^2/12D$, such that we write the time evolution of $g(t)$ as
\begin{equation}
\frac{\partial g\brac{t;\Delta V, \boldsymbol{W}}}{\partial t} = \frac{g_{\infty}\brac{\Delta V,\boldsymbol{W}}-g\brac{t;\Delta V,\boldsymbol{W}}}{\tau},
\label{eq:eqofmotconductance2}
\end{equation}
where it is understood  that $\Delta V$ can be time-dependent (and that $\tau$ depends on $\boldsymbol{W}$). Clearly, the solution to Eq.(\ref{eq:eqofmotconductance}) requires an initial condition. In Ref.\cite{Rene_IonicNeuromorphic} it was shown that the time evolution of Eq.(\ref{eq:eqofmotconductance}) combined with the static conductance $g_\infty$ of Eq.(\ref{ginf}) accounts realistically for the time-dependent conductance of cone-shaped microfluidic channels. Eq.(\ref{eq:eqofmotconductance}) for the conductance $g_{ij}(t)$ of the channel connecting the pair of nodes $\langle ij\rangle$ are of precisely the same form as Eq.(\ref{eq:eqofmotconductance2}). 

\begin{figure*}[t]
    \centering
    \includegraphics[width=2\columnwidth]{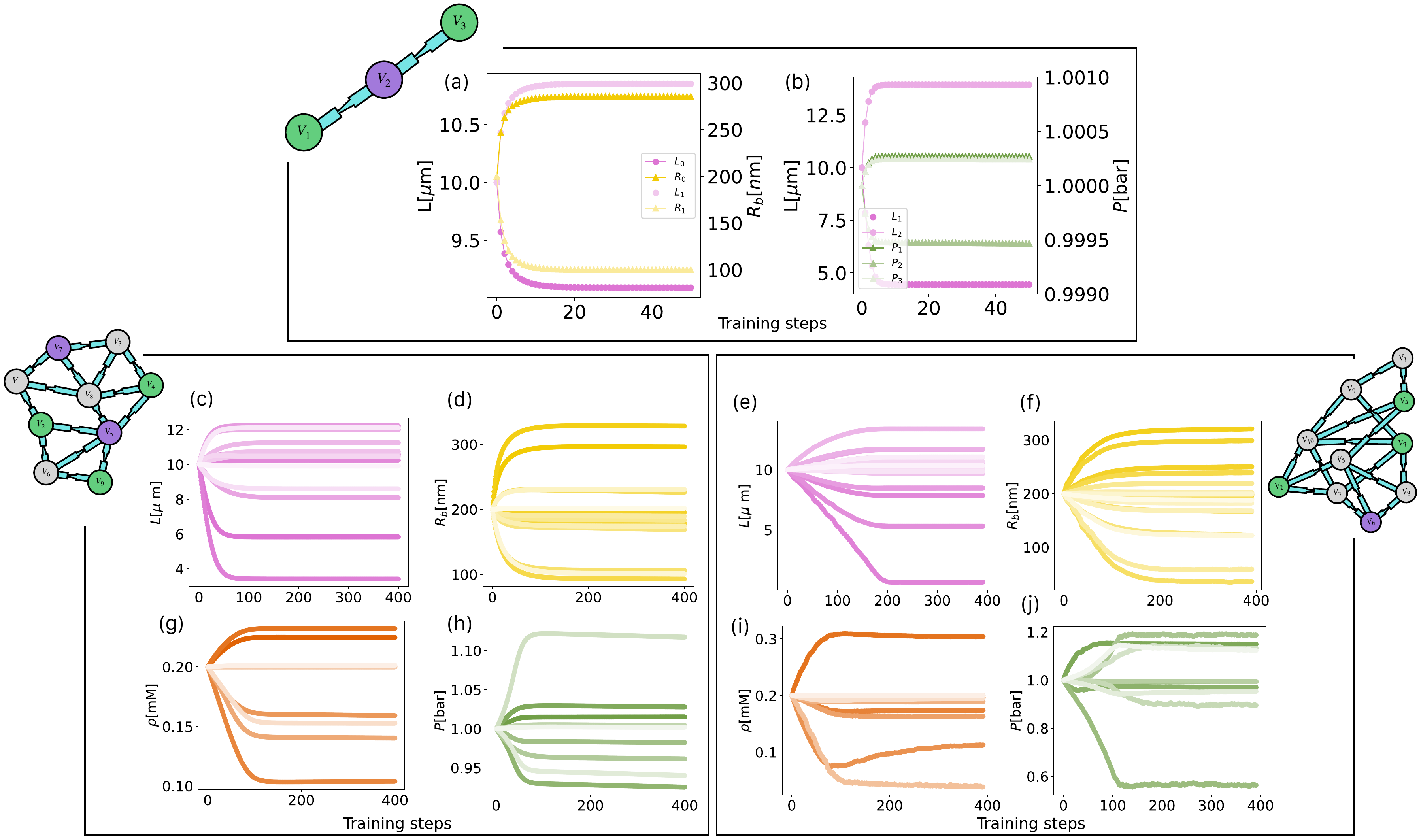}
    \caption{The evolution of the weights during training (a)-(b) Voltage divider trained using different combinations of  features:  (a)  length and base radius, $\boldsymbol{w} = \{L_1, L_2\} \cup \{R_{1}, R_{2}\}$; (b)  length and pressure,  $\boldsymbol{w} = \{L_1, L_2\} \cup \{P_{1}, P_{2}\}$. For a more complex network, weight evolution is shown for an allostery task using (c) length, (d) base radius, (g) ion concentration, and (h) pressure, and for a  linear regression task using (e) length, (f) base radius, (i) ion concentration, and (j) pressure. }
    \label{fig:end_weights}
\end{figure*} 

\subsection{Weight Dynamics}

As training progresses, the weights are iteratively updated to better match the training objectives. In this section, we examine how the weights evolve under different conditions. \cref{fig:end_weights}(a) illustrates the evolution  of the weights for a voltage divider when combining geometrical properties by selecting $\boldsymbol{w} = \{L_1, L_2\} \cup \{R_{1}, R_{2}\}$. \cref{fig:end_weights}(b) presents the changes in weights when combining node- and edge-defined properties by selecting $\boldsymbol{w} = \{L_1, L_2\} \cup \{P_{1}, P_{2}\}$.

For more complex geometries, \cref{fig:end_weights} shows the evolution of the weights for both the allostery and linear regression tasks. We present  the results for the four types of weights  considered: (c) and (e) length, (d) and (f) base radius, (g) and (i) ion concentration, (h) and (j) pressure.

\end{document}